\documentclass[preprint,aps,amsmath,amssymb,12pt,showpacs,showkeys]{revtex4} 

\usepackage{a4c}
\usepackage[dvips]{graphicx}

\newcommand{\n}{\mathbf{n}}
\newcommand{\s}{\mathbf{s}}
\newcommand{\x}{\mathbf{x}}

\renewcommand{\k}{^{(k)}}
\renewcommand{\l}{^{(l)}}

\renewcommand{\ol}[1]{{\overline{#1}}}
\newcommand{\wt}[1]{{\widetilde{#1}}}
\newcommand{\dlangle}{{\langle\!\langle}}
\newcommand{\drangle}{{\rangle\!\rangle_k}}
\def\lsim{\mathrel{\rlap{\lower4pt\hbox{\hskip1pt$\sim$}}
    \raise1pt\hbox{$<$}}}                
\def\gsim{\mathrel{\rlap{\lower4pt\hbox{\hskip1pt$\sim$}}
    \raise1pt\hbox{$>$}}}                
\renewcommand{\section}[1]{{\it #1 --}}

\begin{document}

\addtolength{\voffset}{-\baselineskip}

\title{Nematic order by elastic interactions and cellular rigidity sensing}
\author{B. M. Friedrich, S. A. Safran}
\affiliation{Department of Materials and Interfaces, Weizmann Institute of Science, 76100 Rehovot, Israel}
\date{November 23, 2010}

\begin{abstract}
We predict spontaneous nematic order in an ensemble of active force generators 
with elastic interactions
as a minimal model for early nematic alignment of short stress fibers in non-motile, adhered cells.
Mean-field theory is formally equivalent to Maier-Saupe theory for a nematic liquid. 
However, the elastic interactions are long-ranged (and thus depend on cell shape and matrix elasticity) and originate in cell activity.
Depending on the density of force generators, 
we find two regimes of cellular rigidity sensing for which orientational, nematic order of stress fibers
depends on matrix rigidity either in a step-like manner or with a maximum at an optimal rigidity.
\end{abstract}

\maketitle

\section{Introduction}
The actin cytoskeleton of living cells comprises semiflexible actin filaments and force-generating myosin molecular motors
(as well as many regulatory proteins) \cite{Howard:2001}.
In different cell types, the actin cytoskeleton can display quite distinct degrees of order:
In many motile cell types, the bulk cytoskeleton is simply a cross-linked filament meshwork without any higher degree of ordering,
while in striated muscle cells, actin and myosin assemble into crystal-like myofibril aggregates \cite{Alberts:cell}.
We are interested in an intermediate state of cytoskeletal order found in non-motile, adhered cells, 
where crosslinked bundles of parallelly aligned actin filaments form, so called stress fibers \cite{Deguchi:2009}.
These stress fibers often align with the long axis of the cell,
which establishes nematic order within the ensemble of stress fibers and results in polarized cell forces.
Interestingly, experiments demonstrate a strong influence of substrate rigidity 
on the nematic ordering of stress fibers.
In recent experiments with stem cells, the degree of nematic order of nascent stress fibers
showed a non-monotonic dependence on substrate rigidity
with a maximum at an optimal rigidity \cite{Zemel:2010}.
More generally, substrate rigidity was shown to be a determining factor
for morphology and could even trigger cell fate decisions \cite{Engler:2006}.

In this paper, we address the question of why nascent stress fibers align preferentially with the long
axis of the cell in a substrate dependent manner.
Our work provides a unified theoretical foundation for a previous study, 
which addressed this question in the framework of a linear response theory
that treated the entire cell as a single elastic inclusion \cite{Zemel:2010}.
We predict more general phase behavior for nematic ordering
as a function of substrate rigidity, cell shape anisotropy and contractile force strength.
We focus on the early stages of symmetry breaking by considering a minimal model 
of force generators that is motivated by short, nascent stress fibers, which are coupled to the remaining,
disordered cytoskeleton and which exert contractile forces.
The resulting elastic deformations of the cytoskeleton provide a mean
of long-range communication between distant parts of the cell \cite{Akst:2009}.
We propose that such elastic interactions between nascent stress fibers
drive their nematic alignment with respect to the long axis of the cell
in a way that depends on cell shape and substrate rigidity.

More precisely,
we study the emergence of nematic order within an ensemble of active force generators.
We employ a generic description of the force generators in terms of \textit{extended} force dipoles, 
which are subject to mutual elastic interactions \cite{Wagner:1974,Schwarz:2002}.
The force generators are homogenously distributed within the cellular domain for which we assume suitable boundary conditions.
For simplicity, we model the disordered cytoskeleton as an isotropic, linear elastic material.
While the assumption of isotropy might be well justified for the early stages of cytoskeletal symmetry breaking,
the assumption of a linear elastic material is a strong simplification: 
The cytoskeleton is known to be viscoelastic on long time-scales and may exhibit non-linear elastic behavior for large stresses.
Despite these limitations, 
our minimal model provides qualitative insight into cytoskeletal polarization and reveals different regimes of rigidity sensing.

We find that the cellular shape determines a macroscopic strain field 
due to the long-ranged nature of elastic interactions.
It thus provides a cue for nematic ordering similar to the shape-dependent depolarization factors in electrostatics \cite{Kittel:1986}.
However, in contrast to the electrostatic case where the shape modifies the response to an external electric field, 
the elastic case of active force dipoles shows shape-dependent nematic order even in the absence of external stresses \cite{Zemel:2007}.
Additionally, generic, isotropic hard-core repulsion between force generators
can favor cooperative nematic alignment within the ensemble.
Using a mean field approach inspired by the classical theory of orientational polarization of interacting dipoles \cite{Kittel:1986},
we derive the dependence of nematic ordering on macroscopic boundary conditions 
(such as domain shape and rigidity of a surrounding matrix)
from pairwise elastic interactions.
This dependence provides our prototypical cell with a mechanism to sense and respond to substrate stiffness \cite{Discher:2005}.
We use a mean field approach to predict nematic order of nascent stress fibers as a function of
physical parameters of the surrounding matrix.
In addition to a regime of non-monotonic dependence of nematic order on matrix stiffness with a maximum at
some optimal rigidity,
which had been already found in \cite{Zemel:2010}, we find a second regime, which is characterized 
by a monotonic, step-like dependence. 
Experiments on cells with different aspect ratios observed qualitatively different polarization responses
that match the two regimes of our theoretical model.
Our approach thus generalizes and unifies previous work that either described active cellular responses 
by a phenomenological linear response theory \cite{Zemel:2010},
or considered elastic interactions between individual point force dipoles, but without reference to cell shape or matrix rigidity \cite{Bischofs:2005}.

\section{Extended force dipoles}
We consider a simple model of a mesoscopic force generator embedded in an elastic material,
which we propose as a generic description of 
the tension forces exerted by a short stress fiber on the surrounding, disordered cytoskeleton.
A general force generator located at $\x=0$
exerts a field $f_j(\x)$ of active forces on the cytoskeleton and induces a strain field $u_{ij}(\x)$.
The force field $f_j(\x)$ is characterized by its multipole moments
$P_{i_1,\ldots,i_n,j}=\int d^3\x\, x_{i_1}\cdots x_{i_n} f_j(\x)$ \cite{Schwarz:2002}.
Note that the monopole moment vanishes, $P_i=0$, since the force generator is not acted upon by external forces.
Thus the dipole moment $P_{ij}$ dominates the far field strain.
The far-field strain at distances $|\x|$ much larger than the size $a$ of the 
force generator equals to leading order the strain field $u_{ij}^{\rm far}$ induced by a 
point force dipole with localized force dipole density $p_{ij}(\x)=P_{ij}\,\delta(\x)$ \cite{Schwarz:2002}.
If the cell were infinite,
\begin{equation}
\label{eq_u_far}
u_{ij} \approx u_{ij}^{\rm far} = {\mathcal{G}}_{ik,jl}(\x/|\x|,\nu_c)\,P_{kl} / (E_c |\x|^3)
\end{equation}
where $E_c$ is the Young's modulus of the cell and $\mathcal{G}$ is the angular part of
the force dipole Green's function that depends on its Poisson ratio $\nu_c$ \cite{Landau:elasticity_eng,Schwarz:2002}.
Note that the strain field is long-ranged and decays as $|\x|^{-3}$ just like the electric field of an electric dipole,
but the symmetry of this field expressed by $\mathcal{G}$ is akin to that of an electric quadrupole.
%
The field $u_{ij}^{\rm far}$ does not accurately reflect the strain field 
at small distances comparable to the size $a$ of the force generator.
In particular, $u_{ij}^{\rm far}$ diverges at $\x=0$.
Here, we 
abstract from fine details of the strain field $u_{ij}$ at small distances and propose to 
employ $a$ as a cut-off distance by
introducing a force dipole density $p_{ij}(\x)$ that takes the constant value $P_{ij}/|B|$
inside a spherical region $B=\{|\x|<a\}$ of volume $|B|=4\pi a^3/3$,
but is zero outside $B$. 
Note that the force dipole density $p_{ij}(\x)$ is equivalent to a set of surface forces
$\wt{f}_i=p_{ij}\n_j$ acting on the boundary of $B$, 
where $n_j$ is the surface normal of the region $B$.
This is analogous to the situation in electrostatics, where a uniform electric dipole density
induces the same electrical field as a set of surface charges \cite{Kittel:1986}.

Using Eshelby's classical theory of elastic inclusions, we find that the strain field $u_{ij}^{\rm in}$ 
{\em inside} the spherical region $B$ is uniform \cite{Eshelby:1957,Mura:1991}.
Its purely dilational strain 
$u_{ij}^{{\rm in},0}=u_{kk}^{\rm in}\delta_{ij}/3$ is linear in 
the isotropic part $p_{ij}^{0}=p_{kk}\,\delta_{ij}/3$ of the dipole density,
while the pure shear strain 
$u_{ij}^{\rm in,s}=u_{ij}^{\rm in}-u_{ij}^{\rm in,0}$ 
(which does not involve any volume change)
is linear in the anisotropic part $p_{ij}^{\rm s}=p_{ij}-p_{ij}^{0}$
\begin{equation}
\label{eq_alpha}
u_{ij}^{{\rm in},0}=\alpha_0 p_{ij}^0/E_c\text{ and }
u_{ij}^{\rm in,s}=\alpha_s p_{ij}^s/E_c.
\end{equation}
The coefficients
$\alpha_0=(1/3)(1+\nu_c)/(1-\nu_c)(1-2\nu_c)$ and
$\alpha_s=(2/15)(4-5\nu_c)/(1-\nu_c)(1+\nu_c)$ 
are positive for all physical values of the Poisson ratio $-1<\nu_c<1/2$ \cite{Eshelby:1957}.
For incompressible matrices characterized by $\nu_c=1/2$, the coefficient $\alpha_0$ vanishes.
The strain field $u_{ij}^{\rm out}$ {\em outside} the region $B$ is equivalent to a strain field induced
by the fictitious point force dipole $P_{ij}\delta(\x)$ plus 
some fictitious point force octupole $\wt{P}_{ijkl}\delta(\x)$ located at the center of the ball \cite{Mura:1991}.
We will not require the explicit form of the outer strain field;
for later calculations, it is sufficient to note that its spherical mean $\int_{|\x|=c} d^2\x\,u_{ij}^{{\rm out},k}$ over a
shell concentric with $B$ vanishes.
%
The inner and outer strain field are related by a jump condition that involves the fictitious surfaces force $\wt{f}$ mentioned above,
$u_{ij}^{\rm in}-u_{ij}^{\rm out}(a\n)=3\alpha_0\,\wt{f}_i n_j/E_c$ for $|\n|=1$ \cite{Mura:1991}.
%
%
Previous work considered point force dipoles, which correspond to the limit $a\rightarrow 0$.
Our use of extended force dipoles avoids singularities, 
in particular the elastic deformation energy associated with a force dipole is finite.
Additionally, hard-core repulsion between different force dipoles is introduced in a natural way.
The extended force dipole provides a prototypical example of a force generator and will be used throughout this paper.

\begin{figure}
\begin{center}
\includegraphics[width=10cm]{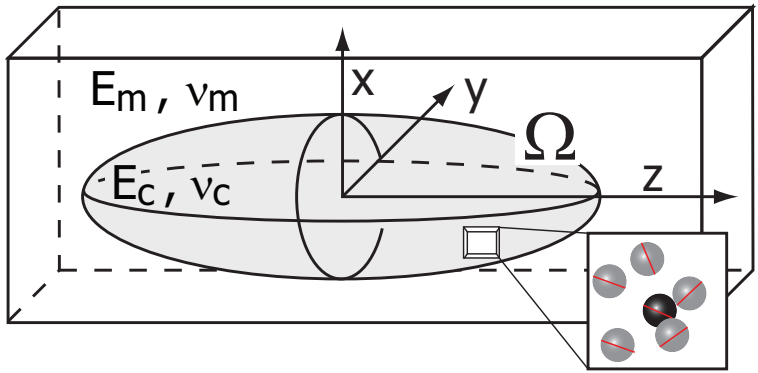}
\end{center}
\caption[]{
\label{fig_cell}
We present a minimal model for the elastic interactions between 
short, nascent stress fibers within a non-motile, adhered cell.
The short stress fibers are modeled as an ensemble of 
force generators that reside inside an elastic cellular domain $\Omega$ 
with Young's modulus $E_c$ and Poisson ratio $\nu_c$.
The cellular domain in turn is embedded in an elastic matrix
(with Young's modulus $E_m$ and Poisson ratio $\nu_m$)
that mimics the substrate to which the cell adheres.
}
\end{figure}

We now consider an \textit{ensemble} of homogenously distributed, extended force dipoles of density $\rho$
that reside inside an elastic domain $\Omega$.
The dipoles have random positions $\x_k$ and dipole moments $P_{ij}\k$.
The individual dipole moments $P_{ij}\k=P_0\, n_i\k n_j\k$ are bipolar
with director $\n_k$. Such a dipole moment equals the dipole moment of a pair of 
opposing point forces $\pm P_0/a\,\n_k$ separated by a vector $2a\n_k$.
The elastic domain $\Omega$ and the force dipoles mimic the 
contractile activity of short, nascent stress fibers embedded into disordered cytoskeleton within an non-motile, adhered cell.
We assume that the domain $\Omega$ is surrounded by an infinitely extended, elastic matrix 
that represents the physical substrate to which the cell adheres, see fig.~\ref{fig_cell}.
The Young's modulus $E_m$ and the Poisson ratio $\nu_m$ of this matrix 
may differ from the respective values $E_c$ and $\nu_c$ of the cellular domain $\Omega$.
To capture the effect of an anisotropic domain geometry, yet keep the calculations simple,
we consider spheroidal domains with semi-axis $a_x=a_y<a_z$
and aspect ratio $r=a_z/a_x$. Thus the $z$-axis denotes the axis of revolution of the cellular domain.
We treat the dipole strength $P_0$ as a control parameter;
experiments suggest that $P_0$ may increase with matrix stiffness \cite{Zemel:2010b}.

We assume that the directors $\n_k$ of the individual force dipoles 
are free to rotate (while their positions $\x_k$ are fixed for simplicity).
In our particular case of cellular domain that has the shape of a prolate spheroid, 
we can show that polarization only occurs along the axis of revolution of the cellular domain.
We can therefore characterize nematic order by a single, scalar order parameter
\begin{equation}
\ol{S}= (3/2)\langle \cos^2\theta_k \rangle_k - (1/2),\quad
\cos\theta_k=\n_{k,z} .
\end{equation}
This order parameter $\ol{S}$ takes the maximal value $1$ for perfect alignment with the $z$-axis
and vanishes for an isotropic director distribution. 
For later use, we also introduce 
the alignment parameters $S_k=(3/2)\cos^2\theta_k-(1/2)$ of all the individual dipoles.

\section{Elastic interactions}
The elastic deformation energy due to the ensemble of extended force dipoles 
is
$H=\int d^3\x\, (\Sigma_l u_{ij}\l)(\Sigma_k\sigma_{ij}\k)/2$
where $\sigma_{ij}\k$ is the elastic stress field induced by the $k$-th dipole \cite{Landau:elasticity_eng}.
Using the 
fact that the force due to an ensemble of dipoles is proportional to the gradient of the dipole density
($\nabla_i \sigma_{ij}\k=\nabla_i p_{ij}\k$),
we rewrite $H$ as \cite{Schwarz:2002}
\begin{equation}
H=\sum_{k,l} U_{kl},\quad
\label{eq_U_int}
U_{kl} = \frac{1}{2} \int_{|\x-\x_k|<a} d^3\s\, u_{ij}^{(l)} p_{ij}^{(k)}.
\end{equation}
Here, $U_{kk}$ represents a ``self-energy'' of the $k$-th dipole%
\footnote{The self-energy $U_{kk}$ diverges, if the spatial cut-off $a\rightarrow 0$;
a UV-cut-off in Fourier space yields a regularized $U_{kk}$ similar to ours, T. Lubensky (private communication).},
whereas $U_{kl}$ represents a pairwise elastic interaction energy
which can be written in the form 
$U_{kl}\sim P_{ij}^{(k)}\mathcal{G}_{im,jn}P_{mn}^{(l)}$
\cite{Wagner:1974,Schwarz:2002}.
It has been proposed that active force generators that are fueled by an external energy reservoir
minimize the work needed to deform the matrix while maintaining a constant 
force magnitude (here $P_0/a$) \cite{Bischofs:2003}%
\footnote{
This is different from passive dipoles, for which their own energetics has to be included \cite{Wagner:1974}.
}.
%
This minimization principle successfully describes the behavior of whole cells adhered to an elastic substrate
in a phenomenological way.
Assuming that cytoskeletal reorganization is governed by local mechanosensitive feedback mechanisms, 
we argue that this minimization principle also applies to sub-cellular cytoskeletal structures such as nascent stress fibers.
This is a crucial assumption of our work and can be tested by comparing our theory to future experiments.


\section{Macroscopic strain fields depend on boundary conditions}
We introduce the macroscopic strain field $\ol{u}_{ij}$ that averages
the microscopic strain field $\Sigma_k u_{ij}\k$ on a length-scale that is much larger than both 
the dipole ball size $a$ and the typical dipole-dipole distance $\rho^{-1/3}$.
Due to the linear relationship between the strain field and the stress sources,
this macroscopic strain field is equivalent to
the strain field induced by the macroscopic force dipole density $\ol{p}_{ij}=\rho \ol{P}_{ij}$ 
where $\ol{P}_{ij}= \langle P_{ij}^{(k)}\rangle$ is the ensemble average of the force dipole moments
\footnote{Unlike \cite{Zemel:2010}, we choose the reference state $\ol{u}_{ij}=0$ for zero 
activity of the force generators with $\ol{p}_{ij}=0$.}.
The force dipole ensemble generates an isotropic macroscopic dipole density 
$\ol{p}_{ij}^0=\ol{p}_{kk}\delta_{ij}/3=\rho P_0 \delta_{ij}/3$
irrespective of nematic ordering.
This ``hydrostatic pressure'' has no analogue in the electrostatic case of electric dipoles,
which have polar as opposed to tensor, nematic symmetry.
A net polarization of the dipole ensemble (here assumed in the $z$-direction) 
gives an additional anisotropic part 
$\ol{p}_{ij}^{\rm s}=\ol{p}_{ij}-\ol{p}_{ij}^0=\rho P_0 \ol{S} (\delta_{iz}\delta_{jz}-\delta_{ij}/3)$ 
which is proportional to the nematic order parameter $\ol{S}$.
For domains of spheroid shape, 
the corresponding macroscopic strain is homogenous throughout the domain $\Omega$ \cite{Eshelby:1957},
similar to the situation in electrostatics \cite{Kittel:1986}.
This homogenous strain $\ol{u}_{ij}$ can be split into a dilation part 
$\ol{u}_{ij}^0=\ol{u}_{kk}\delta_{ij}$
and a pure shear part $\ol{u}_{ij}^{\rm s}=\ol{u}_{ij}-\ol{u}_{ij}^0$. 
Only the shear exerts a torque on a force dipole and is therefore the quantity of interest to us. 
From symmetry considerations, we infer that the shear strain can be written in the form
\begin{equation}
\label{eq_delta}
\ol{u}_{ij}^{\rm s} = (- h+g\ol{S}) \, \rho P_0\, (\delta_{iz}\delta_{jz}-\delta_{ij}/3)/E_c .
\end{equation}
The coefficients $h$ and $g$ can again be computed using Eshelby's theory of elastic inclusions \cite{Eshelby:1957};
they are depicted in fig.~1(a) 
as a function of matrix rigidity $E_m$ for different values of the
domain aspect ratio $r$.
The coefficient $g/E_c$ couples the $z$-anisotropic part of the macroscopic force dipole density
to the macroscopic shear strain $\ol{u}_{ij}^{\rm s}$ and thus provides a macroscopic analogue of the 
shear compliance $1/(2\mu_c)=(1+\nu_c)/E_c$.
The coefficient $h$ is novel in the elastic case and does not exist in the simple electrostatic case; it couples
a ``hydrostatic pressure'' $\ol{p}_{ij}^0$ to a shear strain;
this shape field factor must vanish by symmetry in the case of a spherical domain,
but may be non-zero for non-spherical domains due an anisotropic distribution of elastic restoring forces from
the matrix.
Below, we infer the qualitative behavior of $h$ and $g$ by discussing four important limit cases:
(i) The case of a matrix that is much stiffer than the cellular domain with $E_m\gg E_c$
corresponds to clamped boundary conditions. 
Hence, the macroscopic elastic strain $\ol{u}_{ij}$ vanishes within the cellular domain and $g=h=0$. 
In this case, restoring forces from the matrix completely counterbalance the active cell forces.
(ii) If the matrix is much softer than the cellular domain, $E_m\ll E_c$,
the elastic domain $\Omega$ has essentially stress-free boundary conditions, and there are no restoring forces
from the matrix. 
Hence, the macroscopic elastic stress $\ol{\sigma}_{ij}$ within $\Omega$ equals $\ol{p}_{ij}$.
From $\ol{p}_{ij}^{\rm s}=\ol{\sigma}_{ij}^{\rm s}=2\mu_c \ol{u}_{ij}^{\rm s}$, 
we conclude $g=2\mu_cE_c=1+\nu_c$ and $h=0$, \textit{i.e.} there is no shape-dependence of the macroscopic strain field.
(iii) 
If $\Omega$ is a sphere and has the same elastic properties as the matrix, then
$\ol{u}_{ij}^{\rm s}=\alpha_s\, \rho\ol{P}_{ij}^{\rm s} /E_c$
is computed analogous to the inner strain field of an extended force dipole.
Hence, $g=\alpha_s$, $h=0$.
(iv) Finally, in the limiting case of a prolate spheroid with $r\rightarrow\infty$,
\textit{i.e.} an infinite rod, we have $\ol{u}_{zz}=0$, and hence $h=g >0$.
For a proof, consider the special case 
of a dipole density that is fully polarized in $z$-direction with $\ol{S}=1$.
The strain $\ol{u}_{ij}$ can be equivalently induced by a set
of surface forces $\wt{f}_{i}=\ol{p}_{ij} n_j$.
For a cylinder, $\wt{f}_i=0$ as $n_z=0$.
Thus, $\ol{u}_{ij}=0$ for $\ol{S}=1$, which implies $h-g=0$.
It can be shown that $h>0$ holds true for any prolate spheroid with $0<r\le\infty$.

Consistent with the limiting cases discussed above, $h$ is a non-monotonic function of $E_m/E_c$
that vanishes if the matrix is either very soft or very stiff.
The domain shear compliance $g$ is a monotonically decreasing function of matrix rigidity $E_m$
since the restoring forces, which oppose shear strain $\ol{u}_{ij}^{\rm s}$, increase with $E_m$. 

\begin{figure}
\begin{center}
\includegraphics[width=12cm]{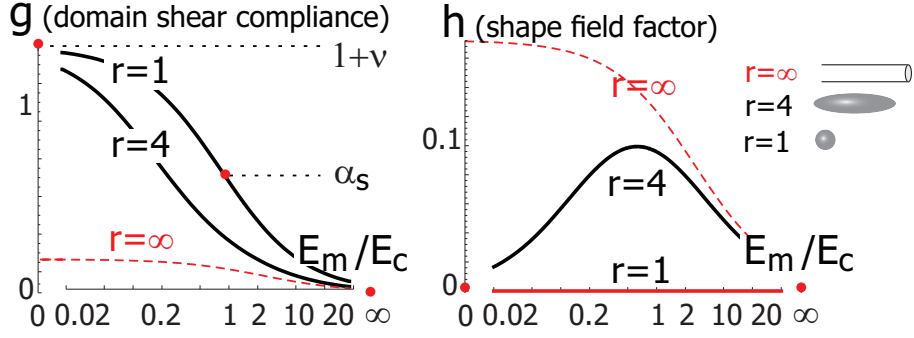}
\end{center}
\caption[]{
\label{fig_g_h}
The macroscopic force dipole density inside the cellular domain ($\ol{p}_{ij}$)
induces shear strain within this domain ($\ol{u}^{\rm s}_{ij}$)
in two different ways, which are characterized by coefficients $g$ and $h$, see eqn.~(\ref{eq_delta}).
Both coefficients depend on the normalized stiffness $E_m/E_c$ of the matrix surrounding the domain.
This is shown for the case of a 
perfectly spherical cellular domain with $r=1$, as well as
for the case of prolate spheroid with aspect ratio $r=4$ and for the case of an infinite cylinder ($r=\infty$).
The ``domain shear compliance'' $g$ characterizes the amount of shear strain, which is due to an anisotropy of the force dipole density itself.
The ``shape field factor'' $h$ characterizes the shear strain that is 
induced 
as a consequence of an asymmetric domain shape and anisotropic restoring forces from the surrounding matrix
by the isotropic ``hydrostatic pressure'' part of the force dipole density.
For the figure, the Poisson ratios are $\nu_c=\nu_m=0.3$;
the limiting cases discussed in the text are marked in red.
}
\end{figure}

\begin{figure}
\begin{center}
\includegraphics[width=8cm]{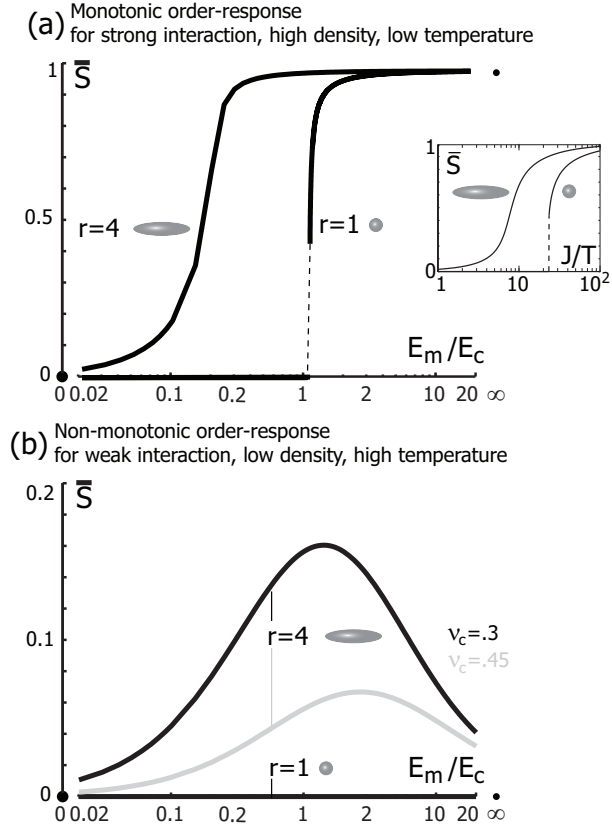}
\end{center}
\caption[]{
\label{fig_Em_S}
Two regimes of rigidity sensing.
The nematic order parameter $\ol{S}$ of the force dipole ensemble 
depends strongly on the normalized stiffness $E_m/E_c$ of the matrix.
Dependent on the effective strength $J/T^\ast$ of their elastic interaction, 
we distinguish two regimes for this dependence:
(a) We observe a monotonic, step-like dependence on $E_m/E_c$ for strong interactions with $J\gg T^\ast$,
\textit{i.e.} for strong dipole strength $P_0$, 
high dipole density $\rho$ and low effective temperature $T^\ast$.
A similar dependence is found for $\ol{S}$ as a function of $J/T^\ast$ if $E_m$ is fixed (here $E_m=2E_c$), see inset.
(b) We observe a non-monotonic dependence with a maximum of nematic order
at an optimal matrix stiffness for weak interactions with $J\ll T^\ast$,
\textit{i.e.} for weak dipole strength $P_0$, 
low dipole density $\rho$ and high effective temperature $T^\ast$.
The Poisson ratios are always $\nu_c=\nu_m=0.3$,
except for panel (b), where we show in gray also the case $\nu_c=0.45$.
}
\end{figure}

\textit{The local strain field} in the vicinity of a typical dipole at position $\x_k$ 
induced by all the other dipoles,
$u_{ij}^{{\rm loc},k}= \sum_{l\neq k} u_{ij}\l$,
is the superposition of the macroscopic strain field $\ol{u}_{ij}$ and a local correction term,
which is similar to a Lorentz cavity field \cite{Kittel:1986}
and insensitive to boundary conditions 
\footnote{
Boundary conditions affect the local correction as terms of order $R^{-2}$,
$R$ is the distance between $\x_k$ and domain boundary;
L. Walpole, Int. J. Eng. Sci. \textbf{34}, 629 (1996).}.
In a continuum approximation that averages over the random positions $\x_l$ of the other dipoles
\footnote{
For regular dipole arrangements \cite{Bischofs:2005},
crystal fields can favor nematic order (\textit{i.e.} for a sc lattice),
but they can also favor anti-ferroelastic order (\textit{i.e.} for a fcc lattice).},
the local strain is induced by a homogeneous dipole density $\ol{p}_{ij}$ 
that fills the entire cellular domain, except for a spherical void around $\x_k$ 
because of hard-core repulsion.
Thus, the local correction approximately amounts to subtracting the strain contribution
of a dipole density $\ol{p}_{ij}$ filling the void
and we obtain for the shear strain
\begin{equation}
\label{eq_u_loc1}
\dlangle u_{ij}^{{\rm loc},k,{\rm s}}(\x)\drangle=\ol{u}_{ij}^{\rm s}-\alpha_{\rm s} \ol{p}_{ij}^{\rm s}/E_c \text{ for }|\x-\x_k|<a,
\end{equation}
By construction, the averaged local strain field looks the same for all force dipoles, 
except possible those which are very close to the boundary of the domain $\Omega$.

\section{Mean field theory for nematic order}
The Hamiltonian $H=\sum_{k,l}U_{kl}$ of pairwise elastic interactions of the force dipoles can be rewritten as
a coupling between their dipole moments and the respective local strain fields
(plus a self-energy term that is independent of the dipole orientations)
\begin{equation}
\begin{split}
H & = \frac{1}{2} \sum_k \int_{|\x-\x_k|<a} \hspace{-0cm} d^3\x\, u_{ij}^{{\rm loc},k} P_{ij}^{(k)}/|B| + U_{kk} .\\ 
\end{split}
\end{equation}
We derive a mean-field Hamiltonian $H_0$ of Lorentz-Weiss type 
from this Hamiltonian 
by replacing the true local strain fields $u_{ij}^{{\rm loc},k}$ 
by their mean field averages $\dlangle u_{ij}^{{\rm loc},k} \drangle$.
Using eq.~(\ref{eq_delta}) (and a standard variational technique, which results in a prefactor 2 \cite{Safran:surfaces}),
we find
\begin{equation}
\begin{split}
\label{eq_U_mean}
H_0 & =
\sum_k \int_{|\x-\x_k|<a} \hspace{-0cm} d^3\x\, \dlangle u_{ij}^{{\rm loc},k} \drangle P_{ij}^{(k)}/|B| \\ 
& = {\rm const.} - \sum_k J \,(\,h - \wt{g}\ol{S}\,)\, S_k  \\
\end{split}
\end{equation}
with $\wt{g}=g-\alpha_s$.
Here $J=2\rho P_0^2/(3E_c)$ 
sets the energy scale of elastic interactions.
This energy scale should be compared to an effective temperature $T^\ast$ 
that quantifies the noise in the system \cite{Zemel:2006}.
Assuming for simplicity Boltzmann statistics for the dipole orientations, 
we self-consistently solve for the nematic order parameter $\ol{S}$ \cite{Wojtowicz:1974}.
In the limit of low interaction strength $|\wt{g}|J\ll T^\ast$,
we find 
$\ol{S} \approx hJ/(5T^\ast+\wt{g}J)$.
For prolate spheroids, the shape field with $h$ is always positive and favors nematic order.
The coefficient $\wt{g}=g-\alpha_{\rm s}$ can be either positive or negative
as the domain shear compliance $g>0$ and the cavity effect of isotropic hard-core repulsion ($\alpha_s$)
give competing contributions to $\wt{g}$.
From eqn.~(\ref{eq_u_loc1}), this can be understood in terms of 
increased or decreased structural interference of 
the strain induced by the ``central'' dipole at $\x_k$
with the macroscopic shear strain $\ol{u}_{ij}^{\rm s}$ 
and the local cavity correction $-\alpha_s \ol{p}_{ij}^{\rm s}/E_c$, respectively.

\begin{figure}
\begin{center}
\includegraphics[width=14cm]{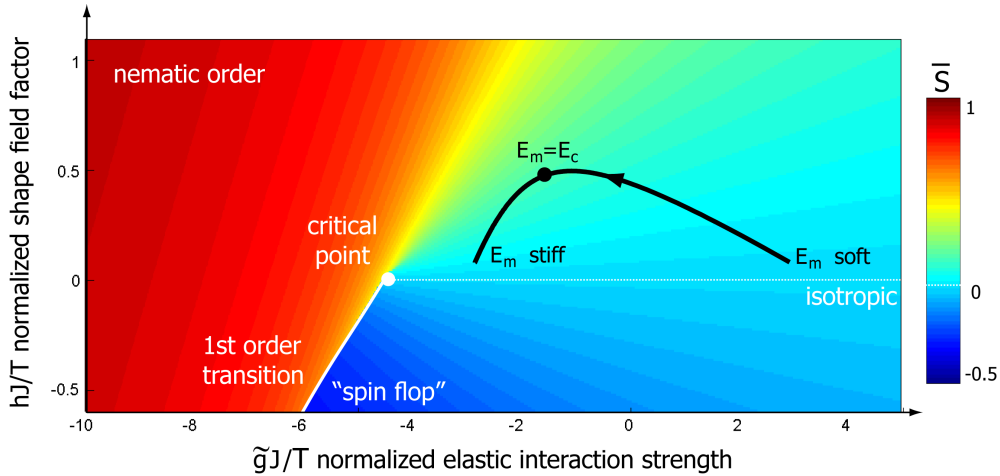}
\end{center}
\caption[]{
\label{fig_phase}
Phase diagram for the nematic order parameter $\ol{S}$ of an ensemble of force dipoles
as a function of the coefficients of the Hamiltonian $H_0$, see eqn.~(\ref{eq_U_mean}).
In our cell model, the coefficients $\wt{g}$ and $h$ depend in turn on domain shape and matrix stiffness in a non-trivial way, see fig.~\ref{fig_g_h}.
As illustration of this dependence, 
we traced out these coefficients 
for an example parameter set for a prolate domain ($r=4$, $\nu_c=\nu_m=0.3$),
which correspond to the black $E_m$-$\ol{S}$ curve in fig.~\ref{fig_Em_S}(b).
}
\end{figure}

The elastic interaction energy of eq.~(\ref{eq_U_mean}) is formally equivalent to the 
generic theory of nematic liquid crystals of Maier and Saupe \cite{Maier:1959} 
with an additional, external alignment field \cite{Wojtowicz:1974}.
In our case, this field arises from the contractility of the system 
(which has no analogue in the electrostatic case) and does not require an external stress.
The field strength $hJ$ is proportional to the shape field factor $h$
and thus depends on the shape of the cellular domain as well as the rigidity of the matrix.
Recall that for $h=0$ (\textit{i.e.} for spherical domains or very stiff or soft matrices),
Maier-Saupe theory predicts a first order phase transition
from an isotropic to a nematic phase when $\wt{g}J/T^\ast\approx -4.542$ \cite{Maier:1959}.
A necessary condition for this transition is $\wt{g}=g-\alpha_{\rm s}<0$,
\textit{i.e.} the cavity effect of isotropic hard-core repulsion has to outweigh
the domain shear compliance $g$.
In the special case of a spherical domain with $r=1$, 
the shape field vanishes, $h=0$, and we have 
$g<\alpha_{\rm s}$ for stiff matrices with $E_m>E_c$, see fig.~\ref{fig_g_h}.
Accordingly, we find strong nematic order for $E_m\gsim E_c$, see fig.~\ref{fig_Em_S}(a).
The full phase diagram exhibits a line of first order phase transition 
that ends in a critical point, see \cite{Wojtowicz:1974} and fig.~\ref{fig_phase}.
Generally, positive values of $h$ increase nematic order and 
smooth out the sharp phase transition that exists when $h=0$, see fig.~\ref{fig_Em_S}(a) for the case $r=4$.
This monotonic, step-like dependence of the nematic order parameter $\ol{S}$ on matrix rigidity $E_m$
is found whenever the elastic interactions are sufficiently strong with $\alpha_sJ/T^\ast>4.542$.
For weak elastic interactions (or large noise strength),
the nematic order parameter $\ol{S}$ 
emulates the behavior of the shape field factor $h$,
and hence depends non-monotonically on matrix rigidity $E_m$ for prolate spheroids, see figs.~\ref{fig_g_h} and \ref{fig_Em_S}(b).
If the cellular domain $\Omega$ is a perfect sphere with $r=1$ and elastic interactions are weak, there is no nematic order.
The results shown are robust with respect to changes of the Poisson ratio $\nu_m$ of the the matrix,
and, in the regime of a monotonic nematic response, also to changes in the Poisson ratio $\nu_c$ of the cellular domain (not shown). 
Only the non-monotonic nematic response is attenuated in nearly incompressible cellular domains and vanishes for $\nu_c=1/2$,
see fig.~\ref{fig_Em_S}(b).
In conclusion, the nematic order of force generators provides a read-out of matrix rigidity
and exhibits qualitatively different regimes of dependency for strong and weak elastic interactions.

It should be noted that our mean field theory is based on a Weiss approximation
and does not correct for a reaction field \cite{Thomas:1965}; 
we thus underestimate $\wt{g}$ and overestimate the transition temperature.
In a real cell, anisotropic hard-core repulsion of actin bundles
can favor nematic order \cite{Onsager:1949} and thus effectively decrease $\wt{g}$.
Both corrections stem from local effects and are insensitive to boundary conditions.

A non-monotonic dependence of cell force polarization on matrix rigidity 
was already predicted in \cite{Zemel:2010} in the framework of a linear response theory,
which assumed $\ol{p}_{ij}^{\rm s} = - \chi_{\rm s} \ol{u}_{ij}^{\rm s}$.
This linear response theory corresponds to the regime of weak elastic interaction $J\ll T^\ast$ in our theory.
Our theory allows a self-consistent derivation of the phenomenological parameter $\chi_{\rm s}$,
which represents an orientational polarization susceptibility: 
In the limit $J\ll T^\ast$, we find $\chi_s=(E_c/5) J/T^\ast$.
Other work studied nematic order by elastic interactions within a two-dimensional
ensemble of force dipoles using Monte-Carlo simulations and found nematic order for high dipole densities \cite{Bischofs:2005}. 
The use of periodic boundary conditions in that study is equivalent to a strain-free boundary 
and should be compared to our limit $E_m\gg E_c$. 

\section{Conclusion}
We showed that elastic interactions can drive nematic ordering of cytoskeletal force generators 
such as short, nascent stress fibers as a function of matrix rigidity.
A non-monotonic dependence of actin cytoskeleton polarity on matrix rigidity has
indeed been found in experiments with embryonic stem cells \cite{Zemel:2010},
which corresponds to our limit of a low density of force generators and large noise strength.
Interestingly, in high aspect ratio cells, nematic order saturates at high matrix stiffness \cite{Zemel:2010};
this may correspond to our prediction of saturation at high values of $J/T^\ast$
since these cells display more developed stress fibers and hence higher contractile forces.
We speculate that different cells might employ different mechanisms of rigidity sensing
by controlling \textit{e.g.} the strength of active cell forces.

\section{Acknowledgement} 
We thank D. Discher, N. Gov, F. Rehfeldt, A. Zemel for stimulating discussions;
this work was supported by the Israel Science Foundation, the Schmidt Minerva Center (SAS), the German Academic Exchange Service (BMF)
and the historic generosity of the Perlman Family Foundation.

\bibliographystyle{unsrt}
\bibliography{../../../../bibliography/cell_mechanics}

\end{document}